\documentclass[aps,pra,nofootinbib,preprint,amsmath,amssymb,floatfix]{revtex4}
\usepackage{color}
\usepackage{graphicx}

\begin{document}

\newcommand{\tc}{\textcolor}
\newcommand{\g}{blue}
\newcommand{\ve}{\varepsilon}
\title{Axion electrodynamics and axionic Casimir effect }         
\author{ Iver Brevik$^1$  }      
\affiliation{$^1$Department of Energy and Process Engineering, Norwegian University of Science and Technology, N-7491 Trondheim, Norway}
\date{\today}          

\begin{abstract}
A general scheme for axion electrodynamics is given, in which a surrounding medium of constant permittivity and permeability is assumed. Then, as an application we give  simple numerical estimates  for the electromagnetic current density produced by the electrically neutral time-dependent axions $a=a(t)$ in a strong magnetic field. As is known, the  assumption $a=a(t)$ is common under  astrophysical conditions. In the third part of the paper, we consider the implications by assuming instead  an axion amplitude $a(z)$ depending on one coordinate $z$ only. If such an axion field is contained within two large metal plates one obtains an axion-generated splitting of the eigenmodes for the dispersion relation.  These modes yield equal, though opposite, contributions to the pressure on the plates.   We calculate the magnitude of the splitting effect,  in a simple one-dimensional model.
\end{abstract}
\maketitle

\bigskip
\section{Introduction}
\label{secintro}

Pseudoscalar axions  of amplitude $a= a(x)$ ($x$ meaning spacetime) are hypothetical particles which are one of the leading candidates  dark matter. If they can to be  found   experimentally, this would mean an important step forward in our understanding of the Universe's composition and development. The axions are believed to be all-pervading, hardly interacting with ordinary matter at all, and they are "cold" in the sense that  they are moving with nonrelativistic velocity, $v \sim 10^{-3}c$. The range of the axion mass $m_a$ is assumed to extend over a few decades of moderate $\mu$eV/c$^2$. These particles may have originated very early in the Universe's history, about the inflationary times. The existence of them was suggested by Helen Quinn and Roberto  Peccei in 1977, in connection with the strong charge-parity (CP) problem in quantum chromodynamics (QCD),  and the subject has since then attracted considerable interest. Some recent references to axion electrodynamics are \cite{sikivie14,lawson19,asztalos04,sikivie03,mcdonald20,chaichian20,zyla20,arza20,carenza20,leroy20,brevik20,qiu17,fukushima19}).

Since the axions are present everywhere, it should be possible to detect them under terrestrial conditions also, at least in principle. In astrophysical contexts, it is common to assume that they are spatially uniform, ${\bf \nabla}a=0$, but vary periodically in
  time with frequency $\omega_a$. A specific  suggestion about how to detect axions on the Earth was presented in Refs.~\cite{sikivie14,lawson19,asztalos04} (the haloscope approach),   looking for the resonance between the natural electromagnetic oscillations in a long plasma cylinder and those from the axion field. A strong magnetic field in the axial $z$ direction was applied. Some extra measures were  necessary, in order  to obtain a cylinder so  'dilute' that it could  make the electromagnetic oscillation frequencies low enough to permit the resonance condition (of order 100 GHz).

  To begin with, we present in the next section a brief overview of the axion-electrodynamic field in the presence of extraneous charges and currents. We allow for a uniform dielectric background, with constant permittivity and permeability. Then, in Sec.~III we give simple numerical estimates  of the axion-generated longitudinal current in the plasma haloscope ~\cite{sikivie14,lawson19,asztalos04} assuming, as mentioned,  that $a=a(t)$. These effects are very small, but actually nontrivial, as the show the existence of electric currents generated by charge-free particles in interaction with a magnetic field.

  In Sec.~IV we consider the opposite extreme, namely an axial field  constant in time but dependent on one spatial coordinate only, $a=a(z)$, in the region between two parallel large metal plates. The dispersion relation shows that there occurs an axion-induced splitting of one of the branches, so that there are two neighboring modes. One mode leads to a weak repulsive Casimir pressure, the other mode reverses the pressure direction. We calculate this effect, making use of scalar electrodynamics in a simple  one-dimensional case.

\section{Basics of axion electrodynamics. Dielectric environment}

The fundamental process is the interaction between a pseudoscalar axion and two photons \cite{mcdonald20}. The Lagrangian describing the electromagnetic field in interaction with the axion field is
\begin{equation}
{\cal{L}}= -\frac{1}{4}F_{\alpha\beta}{H}^{\alpha\beta}  - \frac{1}{4}g_\gamma \frac{\alpha}{\pi}\frac{1}{f_a}a(x) F_{\alpha\beta}\tilde{H}^{\alpha\beta}. \label{1}
\end{equation}
Here $g_\gamma$ is a model-dependent constant of order unity; for definiteness  we adopt the value  $g_\gamma = 0.36$ which follows from the DFS model \cite{dine81,sikivie03}. Further, $\alpha$ is the usual fine structure constant, and $f_a$ is the axion decay constant whose value with is only
insufficiently known; it is often assumed that $ f_a   \sim 10^{12}~$GeV. We assume an isotropic and homogeneous dielectric background, with constant permittivity $\varepsilon$ and permeability $\mu$. When the medium is at rest, the constitutive relations are $\bf{D}=\varepsilon {\bf E}, \ {\bf B}=\mu \bf H$. As is known, in macroscopic electrodynamics there are two field tensors, $F_{\alpha\beta}$ and $H_{\alpha\beta}$, where the latter describes the dielectric response to the fields. We will use the  metric convention $g_{00}= -1$.

 The quantity  multiplying the axion $a(x)$ is thus the product of the electromagnetic field tensor $F_{\alpha\beta}$ and the dual of the response tensor,  $\tilde{H}^{\alpha\beta}=\frac{1}{2}\varepsilon^{\alpha\beta\gamma\delta}H_{\gamma \delta}$, with $\varepsilon^{0123}= 1$. We will use the real metric with $g_{00}=-1$. It is convenient to give the expressions for the field tensors explicitly,
\begin{equation}
F_{\alpha\beta}= \left( \begin{array}{rrrr}
0    &  -E_x   & -E_y  &  -E_z \\
E_x  &    ~0     & B_z   & -B_y  \\
E_y  &  -B_z   &  ~0    &  B_x \\
E_z  &   B_y   &  -B_x &   ~0
\end{array}
\right), \label{2}
\end{equation}

\begin{equation}
{\tilde{H}}^{\alpha \beta} = \left( \begin{array}{rrrr}
0    &  H_x    &  H_y    &  H_z  \\
-H_x &  0      &  -D_z   &  D_y  \\
-H_y &  D_z    &   0     &  -D_x  \\
-H_z &  -D_y   &   D_x   &    0
\end{array}
\right). \label{3}
\end{equation}
Thus $ F_{\alpha\beta}F^{\alpha\beta}= 2(B^2-E^2), \, F_{\alpha\beta}\tilde{H}^{\alpha\beta}= -4\,{\bf E\cdot H}$. The pseudoscalar nature of the interaction is  apparent  from the last expression. The definitions of $F_{\alpha\beta}$ and  $H_{\alpha\beta}$ are covariant; they hold in any inertial system.

With the combined coupling constant $g_{a\gamma\gamma}$ defined as
\begin{equation}
g_{a\gamma\gamma}= g_\gamma \frac{\alpha}{\pi}\frac{1}{f_a}, \label{3a}
\end{equation}
we thus have, for the last term in the Lagrangian (\ref{1}),
\begin{equation}
{\cal{L}}_{a\gamma\gamma} =  g_{a\gamma\gamma} a(x)\,{\bf E\cdot B}. \label{4}
\end{equation}
Based upon the  expression (\ref{1}), the extended Maxwell equations take the following form,
\begin{equation}
{\bf \nabla \cdot D}= \rho-g_{a\gamma\gamma}{\bf H\cdot \nabla}a, \label{5}
\end{equation}
\begin{equation}
{\bf \nabla \times H}= {\bf J}+\dot{\bf D}+g_{a\gamma\gamma}\dot{a}{\bf H}+g_{a\gamma\gamma}{
\bf \nabla}a\times {\bf E}, \label{6}
\end{equation}
\begin{equation}
{\bf \nabla \cdot B}=0, \label{21}
\end{equation}
\begin{equation}
{\bf \nabla \times E} = -\dot{\bf B}. \label{7}
\end{equation}
Here $(\rho, {\bf J})$ are the usual electromagnetic charge and current densities. The equations are so far general; there are   no restrictions on the spacetime variation of $a(x)$. Note again that the equations are covariant, with respect to shift of the inertial system.

\section{Axion-generated electric current in a strong magnetic field}

We now put $\rho = {\bf J}=0$, and consider a geometrical setup essentially being the haloscope model  \cite{sikivie14,lawson19,asztalos04}, whereby a    strong  static magnetic field ${\bf B}_0$ acts in the vertical $z$ direction. The dimension in the $z$ direction is assumed infinite, while  the dimensions in the other directions form a cylinder of radius $R$. It is now natural to employ SI units, whereby the dimension of the axion $a(t)$ becomes J (joule).

 The generalized Maxwell equations given above reduce to their conventional form, except from   Amp\`{e}re's equation which   becomes  modified to
\begin{equation}
{\bf \nabla \times H} =   \frac{g_{a\gamma\gamma}}{c\mu} \dot{a}{\bf B_0}. \label{10}
\end{equation}
Here we have taken into account that  the term containing ${\bf B}_0$ is the dominant term on the right hand side.
The  equation  allows us to  regard the right hand side as an axion-generated electric current density,  ${\bf J}_{\rm axion}$, and  we consider it on the same footing as the ordinary  current density which was called $\bf J$ above.

Now   write the time dependence of the axion as  $a(t)=a_0e^{-i\omega_at}$ with $a_0$ a constant. As mentioned above, the axion velocity is  small, $v/c \sim 10^{-3}$, so that the frequency $\omega_a$ becomes proportional  to the mass, $\hbar \omega_a= m_ac^2$. In our numerical estimates, we will  assume $m_ac^2= 10\, \mu$eV as a typical value. It means that $\omega_a = 1.52\times10^{10}\,$rad/s. This is   a low value, thus justifying the picture of the axion as a classical oscillating field.

As for  the amplitude of $a_0$, we may,   following the notation of Ref.~\cite{duffy09}  express $a(t)$ in terms of the angle $\theta(t)$ characterizing the QCD vacuum state,
\begin{equation}
a(t)= f_a\theta(t). \label{11}
\end{equation}
Taking the axion field to be real, $a(t)=a_0\cos \omega_at$, and similarly $\theta(t)= \theta_0\cos \omega_at$, we have for the amplitudes  $a_0=f_a\theta_0$.  The magnitude of the axion current density  can thus be written as (replacing the permeability with $\mu_0$ for simplicity)
\begin{equation}
{ J}_{\rm axion}(t) =  \frac{g_{a\gamma\gamma}}{c\mu_0} \dot{a}(t){ B}_0=
 - \left( \frac{g_\gamma}{c\mu_0} \frac{\alpha}{\pi} \omega_a {B_0}\right)   \theta_0\sin \omega_at. \label{12}
\end{equation}
Neither the axion amplitude  $a_0$ nor the axion decay constant $f_a$ occur in this expression; the essential quantity being only their ratio $a_0/f_a=\theta_0$.
 Experimental information, such as that coming  from the limits on the electric dipole moment for the neutron \cite{harris99}), indicates that the value of $\theta_0$ is very small. Actually, we may quote the explicit result given in Ref.~\cite{graham11}
 \begin{equation}
 \theta_0 \sim 3\times 10^{-19}. \label{13}
 \end{equation}
 We will here consider  $\theta_0$ as a free parameter, without assigning a numerical value to it.
Inserting the values already mentioned, $g_\gamma= 0.36, \, B_0= 10~$T, \, $\omega_a = 1.52\times10^{10}\,$rad/s, we obtain
\begin{equation}
J_{\rm axion}(t)= -3.37\times 10^5\times \theta_0\sin \omega_at \quad {\rm [A/m^{2}]}.\label{14}
\end{equation}

 Let us go one step further in this direction, by exploiting that the
 local axion energy density is about  0.45 GeV/cm$^3$ \cite{lawson19}. Equating this to $(m_ac^2)N$ with $m_ac^2 = 10~\mu$eV and $N$ the number density of axions, we obtain
\begin{equation}
N= 4.5\times 10^{19}~{\rm m}^{-3}. \label{15}
\end{equation}
This makes it  possible  to introduce a fictitious  effective electric charge $e_{\rm eff}$ per axion. We  can write
\begin{equation}
e_{\rm eff}(Nm_a)v = J_{\rm axion}, \label{16}
\end{equation}
whereby, with $v  \sim 10^{-3}c$, we get the estimate
\begin{equation}
e_{\rm eff} \sim 10^{21}\times \theta_0, \label{17}
\end{equation}
with dimension C (coulomb). This is a physically a huge number, even with $\theta_0=10^{-19}$. Let us therefore recall the background for this calculation: there are reasonable parameters behind the axion current density (\ref{14}), and there is common agreement about the axion energy density being around 0.45 GeV/cm$^3$. The axion number density (\ref{15}) seems also reasonable. It is  thus an open question whether the expression (\ref{17}) has a physical meaning; the very idea of associating axions with a fictitous electric charge may be untenable. For the effective charge to be of the same order of magnitude as the electron charge, the value of  $\theta_0$ would have to be many orders of magnitude smaller than commonly assumed.

\section{Spatially varing axion; Casimir-like  effect}

We will now investigate a typical case where the axion field $a$ is constant in time, but varies with position. For definiteness we adopt the usual geometric setup characteristic for Casimir investigations, namely two large and parallel metal plates separated by a gap $L$. We assume zero temperature. In the region between the plates, we assume that $a(z)$ increases linearly with respect to
the direction $z$ orthogonal to the plates,
\begin{equation}
a(z)= \frac{a_0z}{L}, \quad 0<z<L, \label{18}
\end{equation}
where $a_0$ is the fixed axion value at the plate $z=L$. Outside the plates we assume for definiteness that the values of $a$ are constant, $a=0$ for $z<0$ and $a=a_0$ for $z>a$.

First of all, let us  manipulate the generalized Maxwell equation above to obtain the field equations for the electric and magnetic fields (now in the  Heaviside-Lorentz system of units again),
\begin{equation}
\nabla^2 {\bf E}-\varepsilon\mu \ddot{\bf E}=\frac{1}{\varepsilon}{\bf \nabla}\rho
 +\mu \dot{\bf J}+g_{a\gamma\gamma}\frac{\partial}{\partial t}\left[\dot{a}{\bf B}+\mu {\bf \nabla}a{\bf \times E}\right], \label{19}
\end{equation}
\begin{equation}
\nabla^2 {\bf H}-\varepsilon\mu \ddot{\bf H}= -{\bf \nabla \times J}-g_{a\gamma\gamma}{\bf \nabla \times }[\dot{a}{\bf H}+{\bf \nabla}a{\bf \times E}]. \label{20}
\end{equation}
These equations can be simplified if we omit second order derivatives of the axion, that means time derivatives $\ddot{a}$, space derivatives $\partial_i\partial_j a$, as well as the mixed $\partial_i\dot{a}$. Some manipulations then give us the reduced field equations
\begin{equation}
\nabla^2 {\bf E}-\varepsilon\mu \ddot{\bf E}=  \frac{1}{\varepsilon}{\bf \nabla}\rho +\mu {\dot{ \bf J}}+   g_{a\gamma\gamma}[ \dot{a}\dot{\bf B} +\mu {\bf \nabla}a{\bf \times \dot{E}}], \label{21}
\end{equation}
\begin{equation}
\nabla^2 {\bf H}-\varepsilon\mu \ddot{\bf H} = -{\bf \nabla \times J}-g_{a\gamma\gamma}\left[   \dot{a}{\bf \nabla \times H}  + \frac{\rho}{\varepsilon}{\bf \nabla}a -[({\bf \nabla}a) \cdot {\bf \nabla]  E}      \right] \label{22}
\end{equation}
Now put $\rho = {\bf J}=0$, and observe the condition (\ref{18}) on the axion field. Equations (\ref{21}) and (\ref{22}) reduce to
\begin{equation}
\nabla^2 {\bf E}-\varepsilon\mu \ddot{\bf E}= g_{a\gamma\gamma}\frac{\mu a_0}{L}{\bf \hat{z}\times \dot{E}}, \label{23}
\end{equation}
\begin{equation}
\nabla^2 {\bf H}-\varepsilon\mu \ddot{\bf H} = g_{a\gamma\gamma}\frac{a_0}{L}\partial_z\bf E. \label{24}
\end{equation}
Now going over to Fourier space, with
${\bf E}={\bf E}_0 \exp{[i({\bf k\cdot r}-\omega t)]}, $
 we obtain from Eq.~(\ref{23}) the  component equations
\begin{equation}
(-{\bf k}^2 +\varepsilon \mu \omega^2)E_x-g_{a\gamma\gamma}\frac{\mu a_0}{L}(i\omega)E_y=0, \label{25}
\end{equation}
\begin{equation}
(-{\bf k}^2 +\varepsilon \mu \omega^2)E_y  +g_{a\gamma\gamma}\frac{\mu a_0}{L}(i\omega)E_x=0, \label{26}
\end{equation}
\begin{equation}
(-{\bf k}^2 +\varepsilon \mu \omega^2)E_z = 0. \label{27}
\end{equation}
These equations show that there are two dispersive branches. The first, following from Eq.~(\ref{27}), is the common branch in axion-free electrodynamics,
\begin{equation}
|{\bf k}| = \sqrt{\varepsilon \mu}\,\omega, \quad k_z = \frac{\pi n}{L}, \quad n=1,2,3... \label{28}
\end{equation}
The second branch follows from Eqs.~(\ref{25}) and (\ref{26}) as
\begin{equation}
{\bf k}^2 = \varepsilon \mu \omega^2 \pm g_{a\gamma\gamma}\frac{\mu a_0 \omega}{L}. \label{29}
\end{equation}
This branch is thus composed of two modes, lying very close to the first mode above. For a given $\omega$, there are in all three different values of $|{\bf k}|$. As $g_{a\gamma\gamma}$ is very small, we may replace $\omega$ with $|{\bf k}|/\sqrt{\varepsilon \mu}$ in the last term in the last equation  and solve with respect to $\omega$,
\begin{equation}
\omega = \frac{1}{\sqrt{\varepsilon \mu}}\left[ |{\bf k}| \pm  g_{a\gamma\gamma}\frac{a_0}{2L}\sqrt{\frac{\mu}{\varepsilon}}\right],\label{30}
\end{equation}
neglecting terms of order $g_{a \gamma\gamma}^2$. This  kind of splitting of one of the branches into two slightly separated modes is encountered also in the analogous formalisms given in Refs.~\cite{sikivie03}  and \cite{mcdonald20}.

Let us calculate the zero-point energy $\cal{E}$ of the field, considering the second branch (\ref{30}) only, since this is of main interest. We will consider scalar electrodynamics, meaning that the vector nature of the photons is accounted for, but not their spin. At temperature $T=0$ the energy is $\frac{1}{2}\sum \omega$. We will write the energy in the form
\begin{equation}
{\cal{E}}= \frac{1}{2\sqrt{\varepsilon\mu}}\sum_{n=1}^\infty\left[
\int \frac{d^2 k_\perp}{(2\pi)^2}
  \sqrt{k_\perp^2+\frac{\pi^2n^2}{L^2}}
 \pm \frac{\pi \beta}{L}\right], \label{31}
\end{equation}
where we have defined $\beta$ as
\begin{equation}
\beta = g_{a\gamma\gamma}\frac{a_0}{2\pi}\sqrt{\frac{\mu}{\varepsilon}}. \label{32}
\end{equation}
For the small axion-related part of the energy, we have omitted the continuous part involving ${\bf k}_\perp$.

The first term in the expression (\ref{31}) can be evaluated using dimensional regularization (cf., for instance, Ref.~\cite{milton01}). Replacing the transverse spatial dimension with a general $d$, we can write the first term, called ${\cal{E}}_I$,
 as
\begin{equation}
  {\cal{E}}_I=     \frac{1}{2\sqrt{\varepsilon\mu}}\sum_{n=1}^\infty
\int \frac{d^d k_\perp}{(2\pi)^d}\int_0^\infty \frac{dt}{t}t^{-1/2}\exp\left[  -t\left( k_\perp^2+\frac{\pi^2n^2}{L^2}\right)\right] \frac{1}{\Gamma(-\frac{1}{2})}, \label{33}
\end{equation}
where $\Gamma$ is the gamma function with $\Gamma(-1/2)= -2\sqrt{\pi}$. We have here employed the Schwinger proper time representation of the square root. We integrate over ${\bf k}_\perp$,
\begin{equation}
\int \frac{d^d k_\perp}{(2\pi)^d}e^{-tk_\perp^2} = \frac{t^{-d/2}}{(4\pi)^{d/2}}, \label{34}
\end{equation}
so that
\begin{equation}
{\cal{E}}_I   = -\frac{1}{4\sqrt{\pi\varepsilon\mu}} \, \frac{1}{(4\pi)^{d/2}}\sum_n \int_0^\infty \frac{dt}{t}t^{-1/2-d/2}\exp\left( -\frac{t\pi^2n^2}{L^2}\right). \label{35}
\end{equation}
The sum over $n$ can now be evaluated,
\begin{equation}
  {\cal{E}}_I = -\frac{1}{4\sqrt{\pi \varepsilon\mu}}\frac{1}{(4\pi)^{d/2}}\left( \frac{\pi}{L}\right)^{d+1}\Gamma\left( -\frac{d+1}{2}\right) \zeta(-d-1), \label{36}
\end{equation}
where $\zeta$ is the Riemann zeta function.

We can now take into account the reflection property
\begin{equation}
\Gamma\left(\frac{z}{2}\right)\zeta(z)\pi^{-z/2}= \Gamma\left( \frac{1-z}{2}\right)\zeta(1-z)\pi^{(z-1)/2}, \label{37}
\end{equation}
to get
\begin{equation}
{\cal{E}}_I= -\frac{1}{2^{d+2}\pi^{d/2+1}}\, \frac{1}{\sqrt{\varepsilon\mu}}\, \frac{1}{L^{d+1}}\, \Gamma(1+\frac{d}{2})\zeta (2+d).\label{38}
\end{equation}
Now substituting $d=2$, using that $\zeta(4)=\pi^4/90$, we obtain for the total zero-point energy
\begin{equation}
{\cal{E}} = \frac{1}{\sqrt{\varepsilon\mu}}\left[ -\frac{\pi^2}{1440}\frac{1}{L^3} \pm\frac{\pi \beta}{L}\zeta(0)\right]. \label{39}
\end{equation}
The last term is evidently the small correction from the axions. propagating in the $z$ direction. We will regularize the term simply by using  the analytically continued zeta function, as this recipe has turned out to be effective and correct under the usual physical conditions in spite of lack of mathematical rigor.  Thus we substitute $\zeta(0)=-1/2$, and  obtain
\begin{equation}
{\cal{E}} = \frac{1}{\sqrt{\varepsilon\mu}}\left[ -\frac{\pi^2}{1440}\frac{1}{L^3} \mp  \frac{\pi\beta}{2L} \right].  \label{37}
\end{equation}
This is the total Casimir energy as it is dependent on the gap $L$. The Casimir pressure on the plates follows as $P=- \partial{\cal E}/\partial L$, and is attractive.

What is of main interest, however, is the contribution from particles (photons and axions) moving in the transverse direction $z$. This is what we will call the Casimir energy ${\cal E}_C$. From Eq.~(\ref{31}) we see that this amounts to extracting the terms
\begin{equation}
{\cal E}_C = \frac{1}{2\sqrt{\varepsilon\mu}}\,
\frac{\pi}{L}\sum_{n=0}^\infty (n \pm \beta). \label{41}
\end{equation}
This brings us to the  Hurwitz zeta function, originally defined as
\begin{equation}
\zeta_H(s,a)= \sum_{n=0}^\infty (n+a)^{-s}, \quad (0<a<1, \quad \Re{s}>1). \label{42}
\end{equation}
This function often turns up in Casimir-like problems (cf., for instance, Refs.~\cite{elizalde94,elizalde95,brevik03}). The function has a simple pole at $s=1$. When $\Re s$ differs from unity, the function is analytically continued to the complex plane. For practical purposes one needs only the property
\begin{equation}
\zeta_H(-1,a)=-\frac{1}{2}\left( a^2-a+\frac{1}{6}\right). \label{43}
\end{equation}
Thus we obtain, when omitting the small $\beta^2$ term,
\begin{equation}
{\cal{E}}_C = \frac{1}{4\sqrt{\varepsilon\mu}}\left( -\frac{\pi}{6L} \pm \frac{\pi \beta}{
L}\right). \label{44}
\end{equation}
The first term in this expression comes from the scalar photons propagating in the $z$ direction (it may be noted that the transverse oscillations of a closed uniform string of length $L$ has a Casimir energy of $-\pi/(6L)$; cf. \cite{brevik03}). The second term is the axionic contribution. Recall from Eq.~(\ref{32}) that $\beta$ is independent of $L$. As for the $L$ dependence, the Casimir energies for the one-dimensional  electrodynamic   and the axion parts behave similarly, as one would expect.

In the above equations, the upper and lower signs match each other. Note  that in Eq.~(\ref{44}), the small increase  of the Casimir  energy because of the axions comes from the particular   mode in the dispersion relation (\ref{30}) that is superluminal (meaning that the group velocity is larger than $1/\sqrt{\varepsilon\mu}$). This mode corresponds to a weak repulsive Casimir force. The other mode corresponds to a weak attractive force.

We have examined the two closely separated modes individually.
These modes are physically real, contributing with equal though opposite contributions to the pressure on the plates. In a standard Casimir setup in which only the total pressure is measured, this axionic contribution will thus level out. There might be other cases in the future, however, where these small effects from the modes  could be measurable. The axion-generated eigenmode splitting is definitely of basic physical interest.

\section{Conclusion}

It is notable that current information from astrophysics, implying $a=a(t)$,  indicates that the axions are slowly moving particles in a relativistic sense. In a strong magnetic field, as dealt briefly with in Sec. III, the axions give  rise to a very weak fluctuating electric current, parallel to the magnetic field. From a physical viewpoint this is quite striking, as  an electric current flowing  in a medium with a complex  refractive index necessarily leads to energy dissipation, and in our case the axions are {\it without electric charge}.

  In Sec.~IV, we  investigated the effects  from time-independent but spatially varying  axions in a standard Casimir configuration between two parallel plates. Zero temperature was assumed.  Our formalism was limited to scalar electrodynamics. An important point from a physical viewpoint is the axion-generated splitting of the eigenmodes, resulting in two closely lying modes contributing to the Casimir pressure with equal magnitudes, but of opposite sign. One mode is superluminal corresponding to a weak repulsive  pressure, while the other mode is   subluminal and  corresponds to a weak attractive  pressure.

\end{document}